# High quality factor nitride-based optical cavities: microdisks with embedded GaN/Al(Ga)N quantum dots

M. Mexis,[1] S. Sergent,[2,3] T. Guillet,[1,*] C. Brimont,[1] T. Bretagnon,[1] B. Gil,[1] F. Semond,[2] M. Leroux,[2] D. Néel,[4] S. David,[4] X. Chécoury,[4] and P. Boucaud[4]

[1] *Université Montpellier 2 - CNRS, Laboratoire Charles Coulomb UMR 5221, F-34095, Montpellier, France*

[2] *CHREA-CNRS, 06560, Valbonne, France*

[3] *Université de Nice Sophia Antipolis, Parc Valrose, 06102, Nice, France*

[4] *Institut d'Electronique Fondamentale, CNRS – Université Paris Sud, 91405, Orsay, France*

*Corresponding author: thierry.guillet@univ-montp2.fr

We compare the quality factor values of the whispery gallery modes of microdisks incorporating GaN quantum dots (QDs) grown on AlN and AlGaN barriers by performing room temperature photoluminescence (PL) spectroscopy. The PL measurements show a large number of high Q factor (Q) resonant modes on the whole spectrum which allows us to identify the different radial mode families and to compare them with simulations. We report a considerable improvement of the Q factor which reflect the etching quality and the relatively low cavity loss by inserting QDs into the cavity. GaN/AlN QDs based microdisks show very high Q values (Q›7000) whereas the Q factor is only up to 2000 in microdisks embedding QDs grown on AlGaN barrier layer. We attribute this difference to the lower absorption below bandgap for AlN barrier layers at the energies of our experimental investigation.

III-N semiconductors have attracted much attention for the fabrication of ultra-violet (UV) and visible light sources covering a wide range of applications. Thus, a great effort has been made for realizing low threshold and high performance laser devices operating at room temperature. This has shifted the interest towards new directions such as: i) the fabrication of low-dimensional systems in order to reduce the non-radiative mechanisms generally taking place in bulk materials, especially at high temperature, and ii) the control of their coupling to light through photonic structures aiming at improving the emission and collection efficiencies. Such cavities have previously been successfully implemented for N-based materials and optical modes of high Q values (Q = $\lambda/\delta\lambda$) have been reported up to Q~4000 [1-3]. However, mastering the technology for nitride photonic structures dedicated to UV applications is far more challenging than that of technologically more established materials emitting in infrared for telecommunication applications. This is related to their size which scales down with the wavelength of emission and also to the chemical inertness of III-N materials. Microdisks (µ-disks) are interesting photonic structure for N-based materials [4] sustaining low mode volume and high Q factor Whispering Gallery Modes (WGMs) [5], and also being well adapted to the demonstration of low-threshold laser operation [3] [6] [7].

QDs are promising emitters to be embedded in photonic cavities. With respect to the observation of the photonic modes, an interesting feature of QDs compared to Quantum Wells (QWs) is that they present a larger inhomogeneous broadening. Thus, they allow probing a wide range of cavity modes. Moreover, the optical losses within the cavity can be reduced due the relatively low absorption of light by dots [8]. In addition, QDs have also allowed the observation of phenomena such as strong light-matter coupling [9] and the Purcell effect [8] [10] leading to new low-threshold QD lasers [11] which require only the participation of a small number of dots [12]. For both the blue and UV spectral ranges recent results about GaN/(Al,Ga)N QDs grown on Si have shown that their photoluminescence (PL) remains intense even at room temperature [13] due to the strong quantum confinement of carriers.

In this work we report considerable improvement of the optical quality of µ-disks with embedded GaN/AlN QDs. We investigate the spectroscopy of µ-disks containing GaN QDs grown on two different barrier layers, namely AlN and (Al,Ga)N. By performing micro-PL (µ-PL) measurements we compare the WGMs and their Q factors for both types of barrier layers and various disk diameters. The large number of observed WGMs allows identifying the different radial order families of modes and comparing them with simulations.

The QD samples are grown by ammonia-based molecular beam epitaxy. The main studied sample, shown in figure 1(a) consists of a stack of 4 GaN/AlN QD planes. Each QD plane is formed after the deposition of 7 monolayers (MLs) of GaN. The first QD plane is nucleated on a 35-nm-thick AlN buffer layer grown on a Si (111) substrate. The 4 QD planes are separated by 10-nm-thick AlN spacers and the last plane is capped by a 35-nm-thick AlN layer. The second sample is constituted of a 15 nm AlN buffer layer followed by a 35 nm $Al_{0.5}Ga_{0.5}N$ layer on top of which is grown a GaN QD plane after deposition of 10 MLs of GaN. The QD plane is finally capped by a 50 nm $Al_{0.5}Ga_{0.5}N$ layer.

The limited thickness, i.e. 110 nm, of the nitride based slab for both samples allows us to use the optical active area as a single mode waveguide in the µ-disk plane. Details of the growth, the µ-disks structure and the processing steps can be found in ref [4] [13]. The diameter of the studied µ-disks studied was varied from 2 to 5 µm. The scanning electron microscope (SEM) image in figure 1(b) shows a 2 µm-diameter µ-disk sitting on top of a Si post. The etching quality, deduced from the SEM image, is very high and this is further confirmed by performing optical characterization detailed below.

Our microscopy set-up allows us to photo-excite a single µ-disk with a 266 nm CW laser (Crylas/ FQCW 266-50). The spot diameter is 1.5 µm. The µ-PL was collected both from the top and the edge of the sample –normal and parallel to the micro-disk plane respectively- and was dispersed by a high resolution spectrometer with a spectral resolution of ~0.16 meV. The WGMs are propagating within the µ-disk plane so that the µ-PL collected from the edge show a higher contrast between the modes and the background peak.

In figure 2 we present PL spectra for µ-disk of various radii with embedded GaN/AlN and GaN/(Al,Ga)N dots. For disks incorporating GaN/AlN dots (figure 2(a) and (b)), sharp peaks uniformly and periodically spaced appear across each spectrum, corresponding to WGMs which are distinguishable within a wide spectral range from 2.6 eV to 3.4 eV. We wish to emphasize that such a uniformity of the WGMs across the whole spectrum as well as the fact that these modes are clearly distinguishable for a wide spectral range has not been reported until now in any nitride based µ-disk. Indeed, in the case of µ-disk embedding QWs the modes are clearly observed only at the low energy side of the spectrum due to re-absorption by the QW states at higher energy [7] which limits the Q factor of the modes.

For the microdisks incorporating GaN/AlN dots the Q factor reaches a maximum value of ~5000 for the 2 µm diameter µ-disk and ~7300 for the 5 µm one (inset of figure 2(b)). To our knowledge these Q values are today's state of the art for nitride photonic structures and we attribute this achievement to the high etching quality of the µ-disk and to the low absorption of dots as compared to wells. In the case of

(Al,Ga)N-based µ-disks (figure 2(c)) the Q values are measured smaller by a factor more than 3 than that of AlN ones. The deterioration of the Q factor is attributed to the larger residual absorption of the (Al,Ga)N barriers in the spectral range investigated here, which is not only related to the lower bandgap energy of such barriers but also to the lower crystal quality especially due to alloy disorder [14].

Within the µ-disk the WGMs are described by Bessel functions while the field outside the dielectric is described by Henkel functions. For the Bessel function two mode numbers are introduced in order to describe the eigenmodes; the azimuthal number m and the radial number n [5]. In order to identify the different families of modes (i.e. modes of the same radial number n) from µ-PL spectra we analyze in figure 3 both the Q factor of each mode and the Free Spectral Range (FSR) of each family of modes. Finally, we compare our measurement with the prediction of a theoretical model. The numerical model is based first on the calculation of the guided modes in a standard planar 1-D waveguide [15] and then on solving the Maxwell equations across a boundary of cylindrical symmetry [16]. Due to the thickness of the slab waveguide, typically of a half wavelength in size, only the fundamental order modes are calculated taking into account the spectral dependence of the effective refractive index. The polarization of the observed WGM is predominantly the TE polarization so we have only considered TE modes here.

Figure 3(a) presents the Q values of the sharpest modes that appear within a clear periodic way across the whole µ-PL spectrum of figure 2(a). For the 2 µm µ-disk we assign the different radial mode families by both considering modes with similar Q factor and similar FSR. In figure 3(b) we compare the FSR of these modes to the ones estimated by the model. A good agreement with the model is found for a µ-disk of diameter d=2.06 µm close to the nominal value. From the graph we can distinguish the two first radial mode families, i.e. n=1 and n=2. First order modes (n=1) show the highest Q values because they are propagating closer to the periphery of the disk whereas higher order modes propagating closer to the center of the post. Higher radial order modes, i.e. n>2, are not periodically observed along the PL spectrum because they are damped by the absorption in the Si post. Both the spatial position and the degree of confinement of the first two radial order modes within the disk plane can be observed in figure 3(c) showing an example of the field distribution of simulated WGMs of radial number n=1 and n=2 within the disk plane.

Within the microdisk micro-cavities studied here the limitations of the WGM Q values are described by [6]

$$Q^{-1} = Q_{rad}^{-1} + Q_{scat}^{-1} + Q_{abs}^{-1},  \qquad (1)$$

where the two first factors and are related to photons leaking outside the cavity as photons radiate outside the cavity ( ) or get scattered ( ) due to imperfection that have been induced at the disk periphery during the etching process. The last factor is given by [6]

$$Q_{abs}^{-1} = \frac{\alpha \lambda}{2\pi n_{eff}},  \qquad (2)$$

where is the mean absorption coefficient and is the effective refractive index of the WGM. For both materials the and values which are respectively estimated from the µ-disk geometry and the roughness of the micro-disk sidewalls [16] respectively are calculated to be higher than 10$^4$ through the whole spectral range [13]. Thus the main mechanism responsible for optical losses is absorption ( ) in the silicon post or in the optical active area (wetting layer, QDs and AlN barriers).

We have shown that for the microdisks containing GaN dots grown on AlN layers we can achieve WGM with Q factors of values more than 3 times higher than those with dots grown on AlGaN layer. We attribute this decrease of Q factor for the latter sample to absorption in AlGaN barrier layers. For microdisks embedding GaN/AlN dots we have demonstrated high Q values up to Q~7300. This improvement of the Q value is associated with the etching quality and with the relatively low cavity loss by inserting dots into the cavity. The wide spectral PL emission of QDs at room temperature allows us to identify the first two radial order mode families, n=1 and n=2, by comparing their FSR to simulations.

The authors thank B. Gayral and D. Sam-Giao for their helpful discussion. This work is supported by the French Agence Nationale de la Recherche (research program SINPHONI ANR-08-NANO-021-01).

Figures

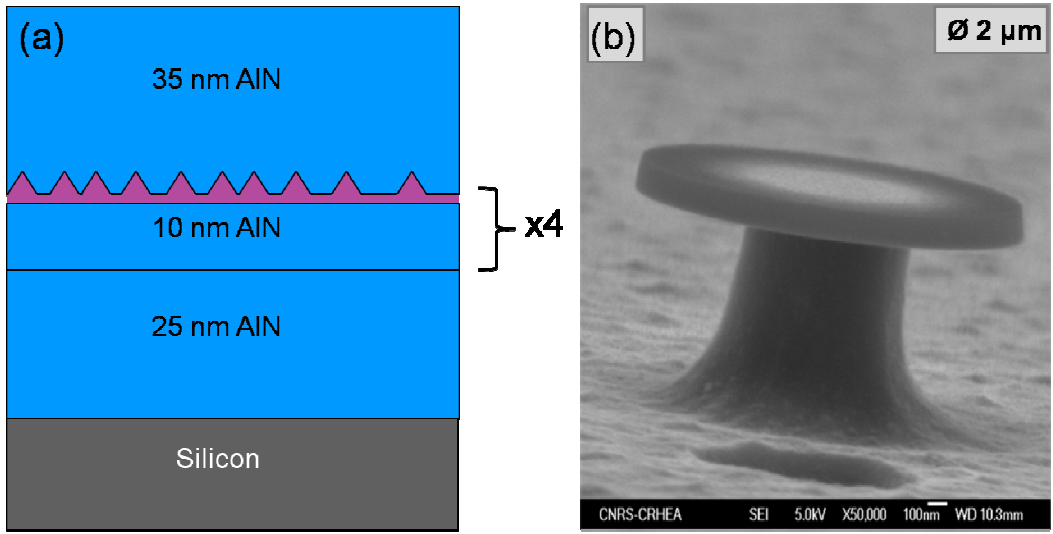

Fig. 1. (color online) a) Epitaxial structure: the optical active area above the Silicon substrate contains 4 periods of GaN/AlN QDs layers. b) SEM image of a 2 µm in diameter µ-disk.

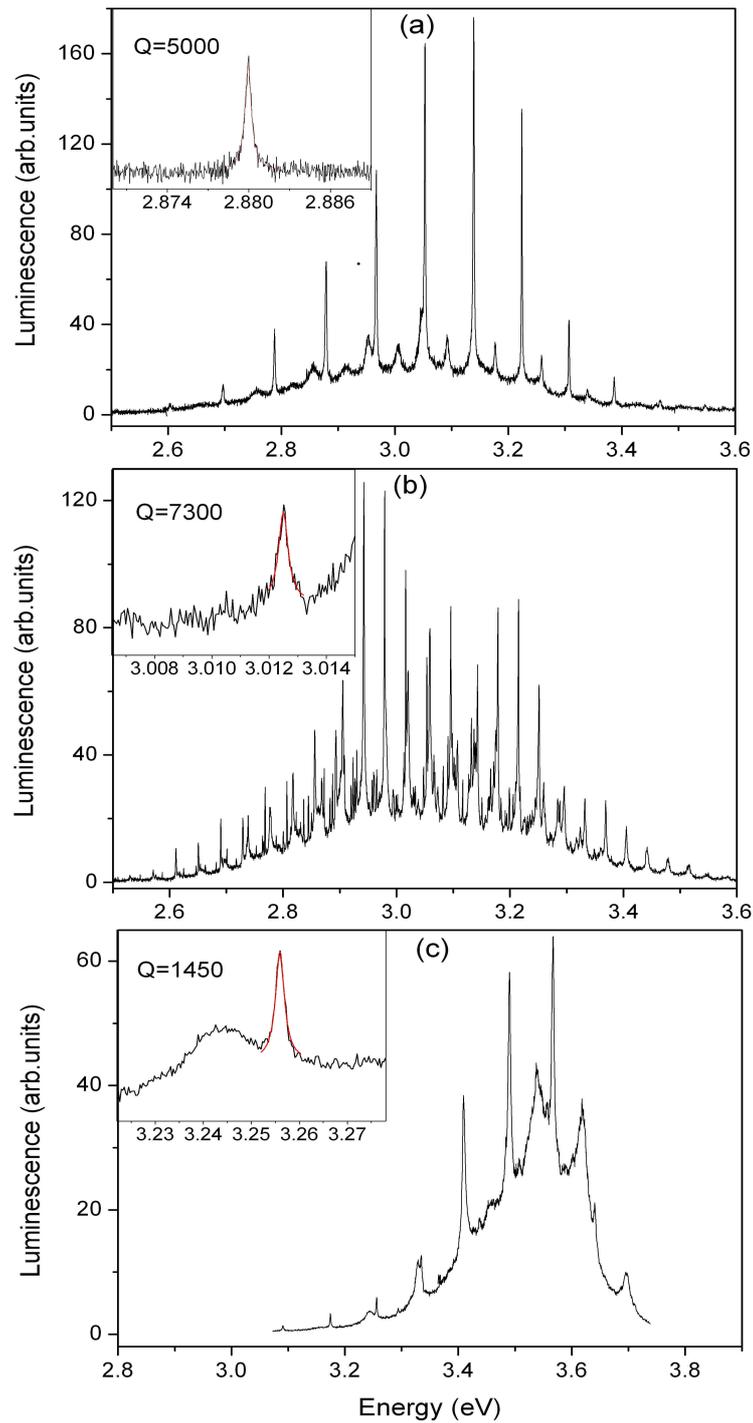

Fig. 2. (color online) Room temperature PL spectra of a),c) 2 μm and b) 5 μm in diameter μ-disks with embedded a),b) GaN/AlN and c) GaN/AlGaN dots presented within similar spectral range ~1.1 eV, under excitation ~60W/cm2. Insets: High resolution spectra showing the WGM with the highest Q value.

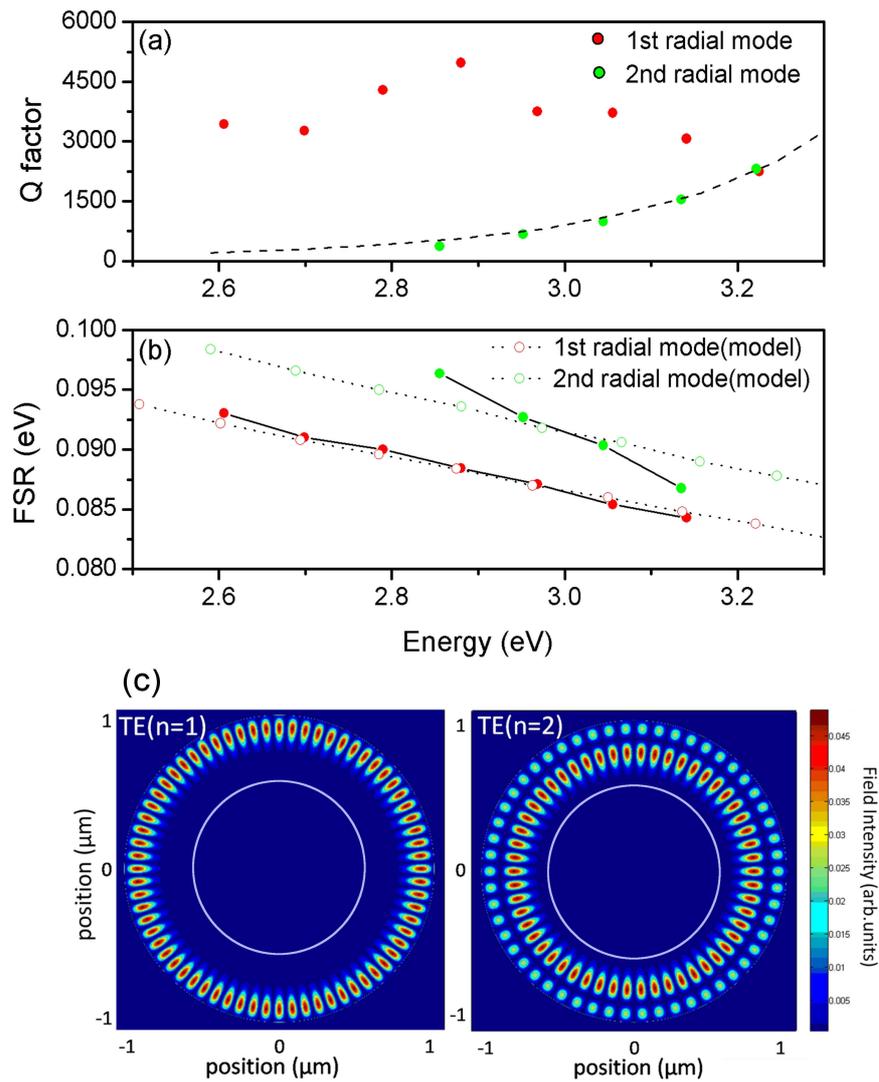

Fig. 3. (color online) The 1st and the 2nd radial order families of WGMs for a ~2 μm μ-disk as these are indicated by a) their Q factor values (the dashed line corresponds to the estimated Q value of the 2nd radial mode when the overlap of the mode wavefunction to the Silicon post is considered) and b) their free spectral range (FSR) (full circles). Theoretical values are shown in open circles. c) An example of simulated 1st (left) and 2nd (right) radial order modes for a 2.06 μm-dimeter μ-disk showing the TE field distribution within the disk plane (the light gray open circle represents the μ-disk post periphery).